\documentclass[twocol]{ametsocV6.1}

\usepackage[utf8]{inputenc}
\usepackage[T1]{fontenc}
\usepackage{anyfontsize} % silence LaTeX Font Warning: Fontshape `OT1...' 
\usepackage{amsmath, amsfonts, amssymb}
\usepackage{xcolor}
\newcommand{\defn}[1]{\emph{#1}}
\renewcommand{\Pr}{\ensuremath{\operatorname{prob}}}

\def\1{\mathbf 1}
\def\p{\boldsymbol\pi}
\def\pct{\%}
\newcommand\restr[2]{{% we make the whole thing an ordinary symbol
  \left.\kern-\nulldelimiterspace % automatically resize the bar with \right
  #1 % the function
  \right|_{#2} % this is the delimiter
  }}

\title{Sampling-Dependent Transition Paths of Iceland--Scotland Overflow Water}

\authors{F.\ J.\ Beron-Vera,\aff{a}\correspondingauthor{F.\ J.\ Beron-Vera,
fberon@miami.edu}
M.\ J.\ Olascoaga,\aff{b}
L.\ Helfmann,\aff{c} 
and P. Miron\aff{d}
}

\affiliation{\aff{a}{Department of Atmospheric Sciencies Rosenstiel School of
Marine, Atmospheric, and Earth Science, University of Miami, Miami,
Florida, USA}\\
\aff{b}{Department of Ocean Sciencies Rosenstiel School of
Marine, Atmospheric, and Earth Science, University of Miami, Miami,
Florida, USA}\\
\aff{b}{Zuse Institute Berlin, Berlin, Germany}\\
\aff{d}{Center for Ocean-Atmospheric
Prediction Studies, Florida State University, Tallahassee,
Florida, USA}
}

\abstract{In this note, we apply Transition Path Theory (TPT) from
Markov chains to shed light on the problem of Iceland--Scotland
Overflow Water (ISOW) equatorward export. A recent analysis of
observed trajectories of submerged floats demanded revision of the
traditional abyssal circulation theory, which postulates that ISOW
should steadily flow along a deep boundary current (DBC) around the
subpolar North Atlantic prior to exiting it. The TPT analyses carried
out here allow to focus the attention on the portions of flow from
the origin of ISOW to the region where ISOW exits the subpolar North
Atlantic and suggest that insufficient sampling may be biasing the
aforementioned demand.  The analyses, appropriately adapted to
represent a continuous input of ISOW, are carried out on three
time-homogeneous Markov chains modeling the ISOW flow. One is
constructed using a high number of simulated trajectories homogeneously
covering the flow domain. The other two use much fewer trajectories
which heterogeneously cover the domain.  The trajectories in the
latter two chains are observed trajectories or simulated trajectories
subsampled at the observed frequency. While the densely sampled
chain supports a well-defined DBC, the more heterogeneously sampled
chains do not, irrespective of whether observed or simulated
trajectories are used. Studying the sampling sensitivity of the
Markov chains, we can give recommendations for enlarging the existing
float dataset to improve the significance of conclusions about
time-asymptotic aspects of the ISOW circulation.}

\begin{document}

\maketitle

\section{Introduction}

The strength of the Atlantic Meridional Overturning Circulation
(AMOC) and its impact on global climate through heat and freshwater
transport is linked to the rates of formation of North Atlantic
Deep Water (NADW) \citep{Buckley-Marshall-16}. Representing the
lower limb of the AMOC, the NADW flows southward.

\begin{figure}[h!]
  \centering%
  \includegraphics[width=\columnwidth]{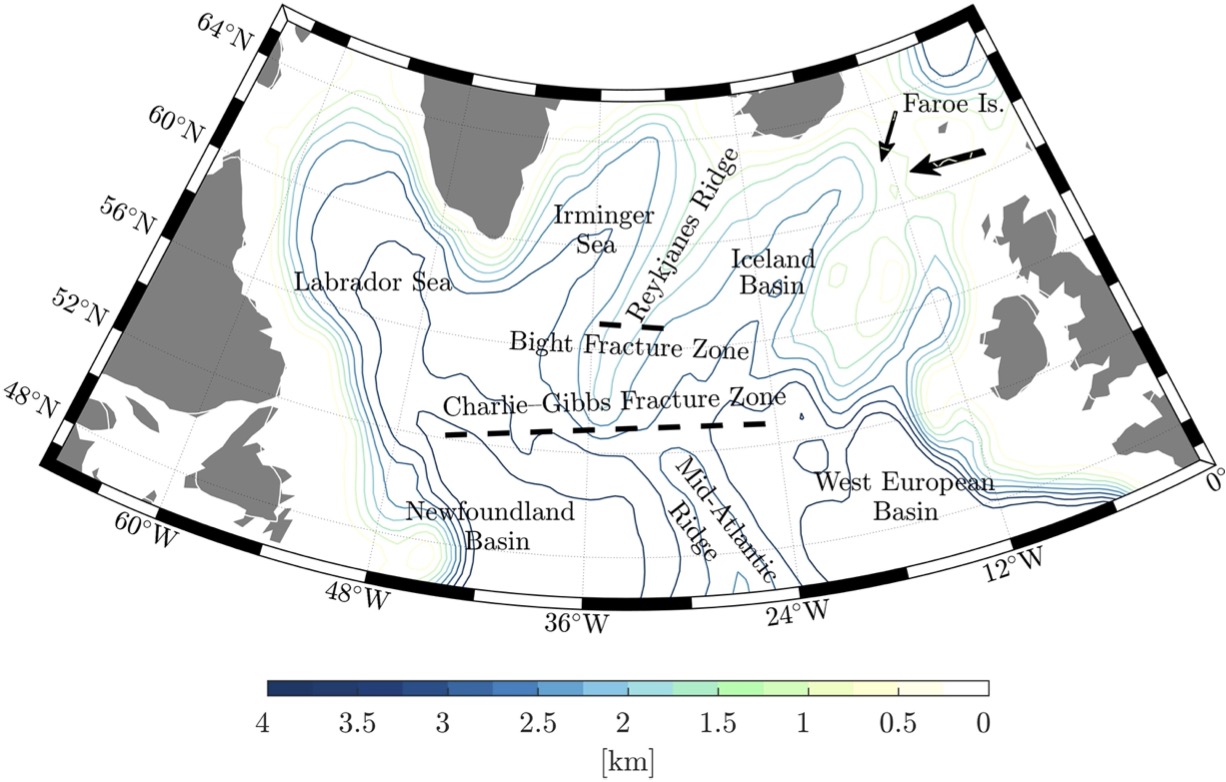}%
  \caption{The subpolar North Atlantic domain of interest with main
  geographic features indicated.  The ISOW enters the Iceland Basin
  primarily through the Faroe Bank Channel (thick arrow) and channels
  across the Iceland--Faroe Ridge (thin arrow). In blue tones,
  selected isobaths.}
  \label{fig:subpolar}%
\end{figure}

The Greenland--Scotland Ridge is a region shallower than 500\,m,
which separates the Nordic seas from the subpolar North Atlantic
(Fig.\@~\ref{fig:subpolar}). The overflow of dense water across
that ridge entrains the thermocline and mid-depth water, representing
one deep-water formation process \citep{Daniault-etal-16}. The
\defn{Iceland--Scotland Overflow Water} (\defn{ISOW}) is the lighter
of the two overflow components of NADW.  It enters the Iceland Basin
mainly through the Faroe Bank Channel and channels across the
Iceland--Faroe Ridge, at a volumetric flow rate of about 5\,Sv
(1\,Sv = 10$^6$ m$^3$\,s$^{-1}$).  The ISOW is characterized by
potential (i.e., with adiabatic heating effects removed) temperature
in the range $-1^{\circ}\text{C} < T_\theta < 2.5^{\circ}\text{C}$
and salinity (salt fraction times $10^{-3}$, approximately) in the
tight interval $34.9 < S < 34.97$.  These $T_\theta$ and $S$ values
roughly correspond to potential sigma-density (density in kg\,m$^{-3}$
minus $10^3$, as is most commonly used in oceanography) in the
narrow range $27.8\text{ kg\,m}^{-3} < \sigma_\theta < 27.9\text{
kg\,m}^{-3}$, which is typically found within 1600--2600\,m.  This
characterization of ISOW follows \citet{Johns-etal-21}, and is
supported by in-situ data collected during the Overturning in the
Subpolar North Atlantic Program (OSNAP) \citep{Lozier-etal-17}.
In-situ observations of the ISOW plume date back to at least the
mid 20th century \citep{Steele-etal-62}; cf.\@~\citet{Johns-etal-21}
for a quite detailed account on the observational efforts that
followed.

The heavier overflow component of NADW is the Denmark Strait Overflow
Water (DSOW), which enters the Irminger Basin between Greenland and
Iceland.  This component is not of our interest here.

The traditional theory of large-scale abyssal circulation of the
ocean postulates that ISOW should steadily flow equatorward, forming
a deep boundary current (DBC) around the subpolar North Atlantic
\citep{Stommel-58}.  The trajectories of 21 acoustically-tracked
isobaric RAFOS (Range and Fixing of Sound) floats \citep{Rossby-etal-86},
deployed at 1800--2800\,m depth as part of OSNAP, challenge this
traditional view, according to \citet{Zou-etal-20}.  By contrast,
they suggest the existence of multiple equatorward ISOW paths, which
represents a puzzle.  A similarly puzzling assessment was made by
\citet{Zou-etal-20} from the inspection of simulated float trajectories
of the same duration (up to 2 years) as, and also longer (10 years)
than, the observed trajectories.  In arriving at their conclusions,
\citet{Zou-etal-20} employed direct inspection of individual observed
float trajectories and the construction of probability distributions
(histograms) of observed and simulated float positions.

The recent application by \citet{Miron-etal-22} of \defn{Transition
Path Theory} (\defn{TPT}) \citep{E-vandenEijnden-06, Metzner-etal-09,
Helfmann-etal-20} to the discretized motion of all available submerged
floats (over 250, of RAFOS and of other type, with the RAFOS floats
corresponding to the totality of those deployed during OSNAP)
represents an attempt to resolve the above puzzle.  TPT highlights
the dominant pathways of fluid parcels connecting a chosen source
with a target region of the flow domain.  That is, instead of
studying the individual complicated paths connecting source with
target, TPT concerns their average behaviour and shows their dominant
transition channels.  Thus TPT leads to a much cleaner picture, and
is hence much easier to interpret than probability distributions
computed using the raw trajectories as in \citet{Zou-etal-20}.
Moreover, the framework for TPT is given by a Markov chain model,
which is constructed from short-run trajectories.  This is advantageous
when dealing with observed trajectories, even when records might
seem too brief to make long-term transport assessments.  Indeed,
under a stationarity assumption, which is appropriate to study
abyssal circulation due to its slow evolution, asymptotic aspects
of the Lagrangian dynamics can be robustly framed by Markov Chains.
By contrast, long trajectory integrations of numerically produced
velocity data, as carried out in \citet{Zou-etal-20}, are sensitively
dependent on initial conditions and hence subject to exponential
divergence with time, making long-term transport assessments dubious.

\citet{Miron-etal-22} found that transition paths of floats can
organize along a DBC, consistent with traditional abyssal circulation
theory.  However, their results are not strictly conclusive for
ISOW as they considered floats in a depth range where the heavier
overflow component of NADW (the DSOW) is also present. 

We revisit the problem of the ISOW export paths from the subpolar
North Atlantic by applying TPT on observed float trajectories
restricted to a depth range within which ISOW is expected to be
better constrained.  We also consider simulated float trajectories,
but along an isopycnic (constant density) layer that intersects the
ISOW density range.  TPT (reviewed in Section 2) is appropriately
adapted (in Section 3) to include the effects of a continuous
injection of ISOW into the subpolar North Atlantic, which represents
a further improvement over \citet{Miron-etal-22}.  This is done by
imposing a mass balance (of ISOW) within the open ocean domain of
interest, which requires one to use a recent adaptation of TPT to
open dynamical systems \citep{Miron-etal-21-Chaos}.  An important
result, among other characterizations of the ISOW problem (cf.\@~Section
4) is that insufficient sampling of the flow domain may mask the
presence of a DBC of ISOW.  Section 5 is dedicated to investigate
the robustness of this result.  The paper is finalized with a summary
and some concluding remarks with recommendations on how to circumvent
the sampling issue (Section 6).

\section{Transition path theory}

Suppose that the long-term motion of fluid parcels can be described
by a stationary stochastic (advection--diffusion) process.  Upon
an appropriate spatiotemporal discretization of the Lagrangian
dynamics, fluid parcels evolve like random walkers of an autonomous,
discrete-time \defn{Markov chain}. That is, conditional on the
current location of the fluid parcel, the Markov chain assigns to
each possible next location a certain transition probability.  Such
a discretization into spatial boxes and discrete time steps can be
achieved using Ulam's method \citep[e.g.,][]{Koltai-10}, by projecting
probability densities onto a finite-dimensional vector space spanned
by indicator functions on \emph{boxes}, which, covering the flow
domain, are normalized by their Lebesgue measure (area)
\citep{Miron-etal-19-Chaos}.  The spatial boxes represent the
locations or states of the chain.

Thus, if $X_n$ denotes the random position at discrete time $nT$,
$n \in \mathbb Z$, on a \emph{closed} two-dimensional flow
domain $\mathcal D$ covered by $N$ disjoint boxes $\{B_1,\dotsc,B_N\}=:
\mathcal D _N$, then $\Pr(X_{n+1}\in B_j) = \sum_{i:B_i\in\mathcal
D_N} P_{ij}\Pr(X_n\in B_i)$ where
\begin{equation}
  P_{ij} := \Pr(X_{n+1}\in B_j \mid X_n \in B_i),\quad
  \sum_{j:B_j\in\mathcal D_N}P_{ij} = 1,
\end{equation}
is the one-step conditional probability of transitioning between
$B_i$ and $B_j$.  The row-stochastic matrix (i.e., with nonnegative
entries and whose rows sum to one)  $\smash{P = (P_{ij})_{i,j :
B_i,B_j\in\mathcal D_N} \in \mathbb R^{N\times N}}$ is called the
\defn{transition matrix} of the Markov chain $\{X_n\}_{n\in  \mathbb{Z}}$.

Let $x(t)$ represent a very long stationary fluid parcel trajectory
visiting every box of the covering of $\mathcal D$ many times. Then
$x_0: = x(t)$ and $x_T: = x(t+T)$ at any $t>0$ provide observations
for $X_n$ and $X_{n+1}$, respectively. These can be used to approximate
$P_{ij}$ via counting transitions between boxes, viz.,
\begin{subequations}
\begin{equation}
  P_{ij} \approx \frac{C_{ij}}{\sum_{l:B_l\in\mathcal D_N} C_{il}}
\end{equation}
where
\begin{equation}
    C_{ij} := \#\big\{x(t)\in B_i,\, x(t+T)\in B_j,\, t:\text{any}\big\}.
\end{equation}
\label{eq:P_estimation}
\end{subequations}

If $P$ is \defn{irreducible} or \defn{ergodic} (i.e., all states
in the Markov chain communicate) and \defn{aperiodic} or \defn{mixing}
(i.e., no state is revisited cyclically), its dominant \emph{left}
eigenvector, $\p$, satisfies $\p P = \p$, and can be chosen
componentwise positive. Scaled appropriately it represents a
(limiting, invariant) \defn{stationary distribution}, namely, $\p
= \smash{\lim_{k\uparrow \infty} \mathbf f P^k}$ for any probability
vector $\mathbf f$ \citep{Norris-98}.  We will assume that the
Markov chain is in stationarity, meaning that $\Pr(X_n \in B_i) =
\pi_i$ for all $n \in \mathbb Z$.

Transition Path Theory (TPT) provides a rigorous characterization
of the ensemble of trajectory pieces, which, \emph{flowing out last
from a region $\mathcal A \subset \mathcal D_N$, next go to a region
$\mathcal B \subset \mathcal D_N$, disconnected from $\mathcal A$}.
Such trajectory pieces are called \defn{reactive trajectories} due
to tradition originated in chemistry, which identifies $\mathcal
A$ resp.\ $\mathcal B$ with the reactant resp.\ product of a chemical
reaction.  In fluidic terms, reactive trajectories most effectively
contribute to the transport between $\mathcal A$ and $\mathcal B$,
which will herein be referred to as \defn{source} and \defn{target},
respectively.

The main objects of TPT are the \defn{forward  committor probability},
$\smash{\mathbf q^+ = (q^+_i)_{i:B_i\in\mathcal D_N}}$, giving the
probability of a trajectory initially in box $B_i$ to first enter
the target $\mathcal B$ and not the source $\mathcal A$,  and the
\defn{backward committor probability}, $\smash{\mathbf q^- =
(q^-_i)_{i:B_i\in\mathcal D_N}}$, giving the probability of a
trajectory in box $B_i$ to have last exit the source $\mathcal A$
and not the target $\mathcal B$.  The committors are fully computable
from $P$ and $\p$, according to \citep{Metzner-etal-09, Helfmann-etal-20}:
\begin{equation}
  \left\{
   \begin{aligned}
   &\restr{\mathbf q^\pm}{\mathcal D_N \setminus (\mathcal A\cup \mathcal B)} =  \restr{P^\pm}{\mathcal D_N \setminus \mathcal A\cup \mathcal B,\mathcal D_N} \mathbf q^\pm,\\
   &\restr{\mathbf q^+}{\mathcal A}  = \mathbf 0^{|\mathcal A|\times 1},\quad \restr{\mathbf q^-}{\mathcal B} = \mathbf 0^{|\mathcal B|\times 1},,\\
   &\restr{\mathbf q^+}{\mathcal B}  = \mathbf 1^{|\mathcal B|\times 1},\quad \restr{\mathbf q^-}{\mathcal A} = \mathbf 1^{|\mathcal A|\times 1}.
  \end{aligned}
  \right.
  \label{eq:q}
\end{equation}
Here, $\restr{}{\mathcal S}$ denotes restriction to the set $\mathcal S$ while $\restr{}{\mathcal S,\mathcal S'}$ that to rows corresponding to $\mathcal S$ and columns to $\mathcal S'$; $P^+ = P$; and 
\begin{equation}
  P^-_{ij} := \Pr(X_n\in B_j\mid X_{n+1}\in B_i) = \frac{\pi_j}{\pi_i}P_{ji}
\end{equation}
are the entries of the \defn{time-reversed transition matrix}, i.e.,
for the original chain traversed in backward time, $\{X_{-n}\}_{n\in\mathbb Z}$.

The committor probabilities are used to express several statistics
of the ensemble of reactive trajectories as follows
\citep{Helfmann-etal-20}:
\begin{itemize}
  \item The \defn{reactive distribution},
	 $\smash{\boldsymbol \pi^{\mathcal A\mathcal B} = (\pi^{\mathcal
	 A\mathcal B}_i)_{i: B_i\in \mathcal D_N}}$, where $\pi^{\mathcal
	 A\mathcal B}_i$ is defined as the joint probability that
	 a trajectory is in box $B_i$ while transitioning from
	 $\mathcal A$ to $\mathcal B$ and is computable as
    \begin{equation}
    \pi^{\mathcal A\mathcal B}_i = q^-_i\pi_iq^+_i.
    \end{equation} 
 \item The \defn{reactive currents},
	$\smash{f^+ = (f^+_{ij})_{i,j : B_i,B_j\in\mathcal D_N}}$,
	where $f^+_{ij}$ gives the net flux of trajectories going
	through $B_i$ at time $nT$ and $B_j$ at time $(n+1)T$ on
	their way from $\mathcal A$ to $\mathcal B$, indicates the
    dominant transition channels. This is computable according
	to
   \begin{equation}
	f^+_{ij} := \max\left\{f^{\mathcal A\mathcal B}_{ij} -
   f^{\mathcal A\mathcal B}_{ji},0\right\},\quad
	f^{\mathcal A\mathcal B}_{ij} = q^-_i \pi_i P_{ij}q^+_j.
	\end{equation}
 \item The \defn{reactive rate} of trajectories leaving
	$\mathcal A$ resp.\ entering $\mathcal
   B$, defined as the probability per time step of a reactive
   trajectory to leave $\mathcal A$ resp.\ enter $\mathcal B$, is
   computed as
   \begin{equation}
	k^{\mathcal A\to\mathcal B} := \sum_{\substack{i:B_i\in\mathcal
   A,\\ j : B_j\in \mathcal D_N}} f^{\mathcal A\mathcal
	B}_{ij} \text{ resp.\ }
	k^{\mathcal B\gets\mathcal A} := \sum_{\substack{i:B_i\in\mathcal
   D_N,\\ j : B_j\in \mathcal B}} f^{\mathcal A\mathcal
	B}_{ij},
   \end{equation}
   and gives the proportion of reactive trajectories leaving
	$\mathcal A$ resp.\ entering $\mathcal B$.  It turns out
	that
	\begin{equation}
	k^{\mathcal A\to\mathcal B} \equiv k^{\mathcal B\gets\mathcal
	A} =: k^{\mathcal A\mathcal B}.
	\end{equation}
	In some situations, as we consider below, it is insightful
	to further decompose the transition rate
	\begin{equation}
      k^{\mathcal B\gets\mathcal A} = \sum_{j : B_j\in\mathcal B}
      k^{B_j\gets\mathcal A}
    \end{equation}
    into the individual arrival rates into the disjoint boxes $B_j$
    that cover $\mathcal B$:
    \begin{equation}
      k^{B_j\gets\mathcal A} = \sum_{i:B_i\in \mathcal D_N} f^{AB}_{ij}.
    \label{eq:decomp_rate}
    \end{equation}
    The same can be done for decomposing $k^{\mathcal A\to\mathcal
    B}$.
 \item Finally, the \defn{reaction duration}, $t^{\mathcal A\mathcal
   B}$, of a transition from $\mathcal A$ to $\mathcal B$ is obtained
   by dividing the probability of being reactive by the transition
   rate interpreted as a frequency, viz.,
   \begin{equation}
    t^{\mathcal A\mathcal B} := \frac{\sum_{i : B_i \in \mathcal D_N}
	\pi^{\mathcal A\mathcal B}_i}{k^{\mathcal A\mathcal B}}.
   \end{equation}
\end{itemize}

\section{The ISOW problem}

For the ISOW problem,  we are interested in the source set given
by the region where ISOW originates and a target set given by the
area where ISOW exits the subpolar North Atlantic. Since the domain
$\mathcal D$ here represents a subdomain of the subpolar North
Atlantic, and thus an \emph{open} flow domain, $P$ cannot be
row-stochastic, which requires an adaptation of TPT.  This is
proposed in \citet{Miron-etal-21-Chaos}, which we further adapt
here to account for the input of ISOW into $\mathcal D$.

Specifically, we replace $P$ by a row-stochastic transition matrix
$\tilde P\in \mathbb R^ {(N+1)\times (N+1)}$ defined by
\begin{equation}
 \tilde P :=
 \begin{pmatrix}
	 P & P^{\mathcal D_N\to\omega}\\ 
	 P^{\mathcal A\gets\omega} & 0
 \end{pmatrix}
 \label{eq:closure}
\end{equation}
on the extended domain $\mathcal D_N \cup \omega.$ Here, $\omega$
denotes a state, called a \defn{two-way nirvana state}, which is
added to the chain defined by $P$. It absorbs probability imbalance
from $\mathcal D_N$, which is sent back to the chain through the
source $\mathcal A$.  More precisely, in \eqref{eq:closure},
\begin{equation}
  P^{\mathcal D_N\to\omega} = \Big(1 - \sum\nolimits_{j : B_j\in
  \mathcal D_N} P_{ij}\Big)_{i : B_i\in \mathcal D_N}\in \mathbb
  R^{N \times 1}
\end{equation}
gives the outflow from $\mathcal D_N$ and 
\begin{equation}
  P^{\mathcal A\gets\omega} = \frac{\1_{\mathcal A}}{|\mathcal
  A|}\in \mathbb R^{1\times N}
\end{equation}
is a probability vector that gives the inflow into $\mathcal D_N$,
taken to take place uniformly through $\mathcal A$, thereby enforcing a mass
balance for ISOW.  In other words, \emph{this guarantees a continuous
inflow of ISOW through $\mathcal A$, whose transition paths into
the target $\mathcal B$, to be placed at the southern edge of the
subpolar North Atlantic (sub)domain  $\mathcal D_N$, can be
unequivocally tracked.}

TPT is adapted in \citet{Miron-etal-21-Chaos} such that transitions
between $\mathcal A$ and $\mathcal B$ are constrained to take place
within $\mathcal D_N$, i.e., they avoid the artificial state $\omega$.
This is shown in \citet{Miron-etal-21-Chaos} to be accomplished by
restricting the committor problem to $\mathcal D_N$. The restricted
committor equations are given by \eqref{eq:q} with the substochastic
matrix $P^+_{ij}=P_{ij}$ on the open domain $\mathcal D_N$ and the
replacement $P^-_{ij} =  \smash{\frac{\tilde\pi_j}{\tilde\pi_i}
P_{ji}}$ for $i,j: B_i,B_j\in \mathcal D_N$.  That is, the transition
matrix for the open time-reversed Markov chain is computed using
the restriction to $\mathcal D_N$ of $\tilde\p$, the stationary
distribution of the \emph{closed} transition matrix $\tilde P$.
Since the probability imbalance enters the open domain via $\mathcal
A$, the transition paths that avoid $\omega$ are unaffected by this
artificial closure of the system.  The rest of the TPT formulae
remain the same, except that for computing the statistics, $\p$ is
replaced by $\restr{\tilde\p}{\mathcal D_N}$.

\begin{figure}[h!]
  \centering%
  \includegraphics[width=\columnwidth]{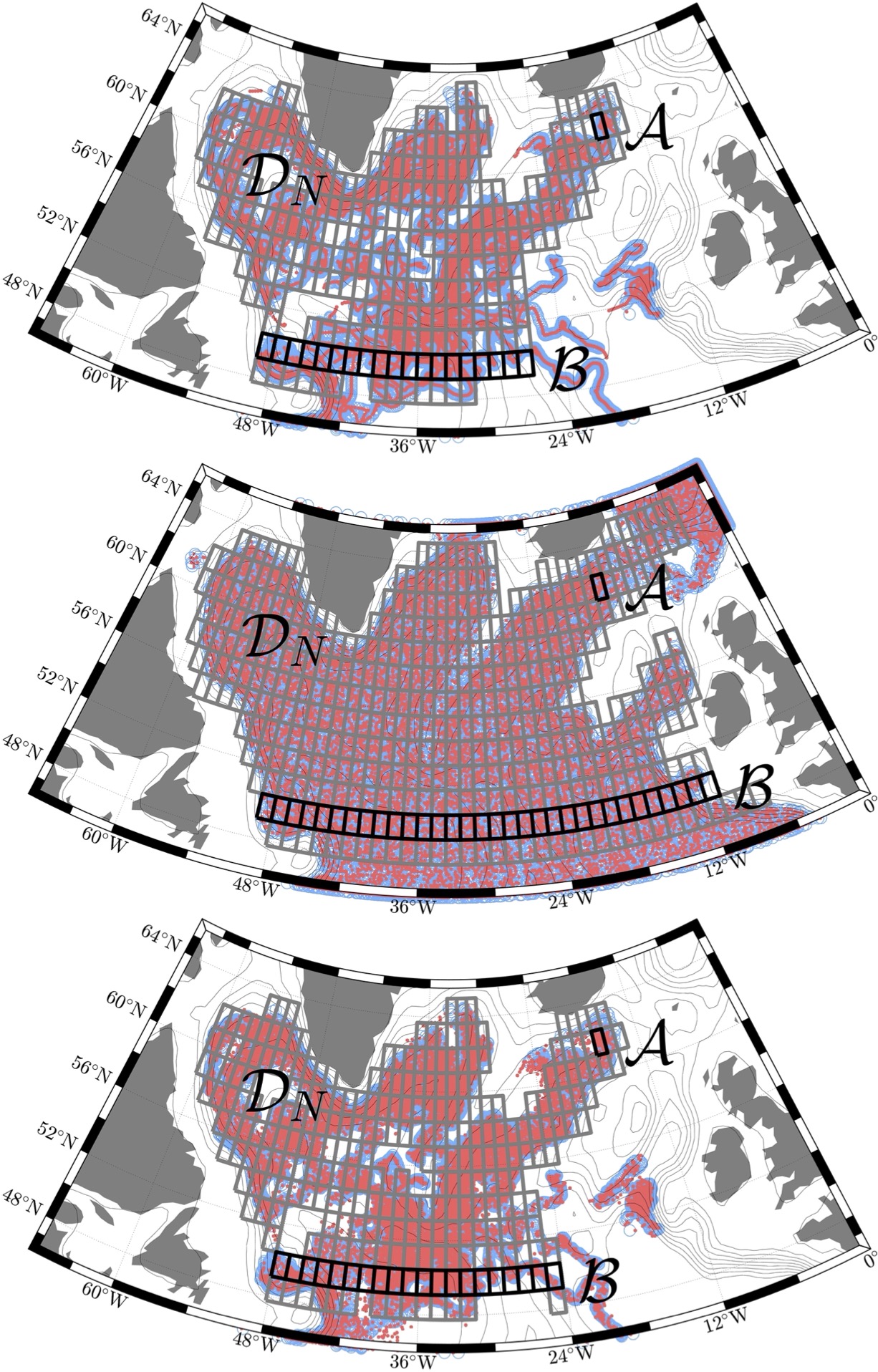}%
  \caption{(top panel) Ignoring the start date, initial positions $x_0$
  (light blue) and positions after 10 days $x_T$ (light red) of available
  acoustically-tracked RAFOS and satellite-tracked profiling Argo
  floats parked between 1600 and 2600\,m, where ISOW is typically
  found. In thick gray, boxes of the domain covering ($\mathcal
  D_N$)  used to construct a maximally communicating Markov chain
  for TPT analysis.  In black, source ($\mathcal A$) and target
  ($\mathcal B$) used in the TPT analysis. The thin gray curves are selected
  isobaths.  (middle panel) As in the top panel, but based on
  simulated float trajectories integrated from HYCOM model velocities
  on an density layer intersecting the typical ISOW density range.
  Unlike in the top panel, the initial positions $x_0$ uniformly cover
  the flow domain.  (bottom panel) As in the middle panel, but only
  using initial positions $x_0$ that lie in closest proximity to those
  of the observed floats.}% 
  \label{fig:D}
\end{figure}

To carry out the TPT analysis of the ISOW problem just posed, we
will consider two datasets for estimating the transition matrix as
in \eqref{eq:P_estimation}.  On the one hand, we have \emph{observed}
trajectories produced by acoustically-tracked RAFOS floats (from
OSNAP and earlier experiments) and satellite-tracked profiling Argo
floats \citep{Argo-00}, deployed in the subpolar North Atlantic
domain of interest (in the 1990s and mid-2010s), or that simply
happened to travel through it (since the 1980s) \citep[cf.][for
details of the two types of isobaric floats]{Miron-etal-22}.  Unlike
in the TPT analysis of \citet{Miron-etal-22}, we restrict to
trajectories parked between 1600 and 2600\,m, a depth range where
ISOW is expected to be found \citep{Zou-etal-20, Johns-etal-21}.
There is a total of 268 float trajectories, with positions interpolated
10-daily. If a given trajectory contains a gap larger than 10 days,
then it is split into two separate trajectories. The transition
time choice of $T = 10$ days and the box size of (roughly)
1$^{\circ}$-by-1$^{\circ}$ were found sufficient to allow maximal
communication by the application of the Tarjan's algorithm
\citep{Tarjan-72} on a time-homogeneous (as we ignore the start day
of the trajectories) Markov chain on boxes over $\mathcal D$ that
covers the largest portion of the subpolar North Atlantic
(Fig.\@~\ref{fig:D}, top panel).  We will refer to this chain as
the \emph{observed chain}.

On the other hand, we consider \emph{simulated} float trajectories
obtained by integrating velocities produced by a 1/12$^{\circ}$
Atlantic Ocean simulation based on HYCOM (HYbrid-Coordinate Ocean
Model), as described in \citet{Xu-etal-13} and considered by
\citet{Zou-etal-20}.  The velocities were extracted (except north
of the domain of interest, where interpolation was involved) on the
model's native isopycnic-coordinate level (layer) $k = 26$, which
intersects the ISOW typical density range. The trajectories were
initialized every 5 days for 2 years from a very dense grid over
the subpolar North Atlantic domain, each one lasting 60 days.
Ignoring the start day and using the same transition time ($T = 10$
days) as with the observed trajectories, a maximally communicating
Markov chain on boxes (of the same size, 1$^{\circ}$-by-1$^{\circ}$,
as in the case of observed trajectories) over $\mathcal D$ is
obtained, which covers the subpolar North Atlantic as shown in the
middle panel of Fig.\@~\ref{fig:D}.  Unlike in the observed trajectory
case, in which there are on average about 100 data points per box,
the simulated trajectory case contains on average 1500 data points
per box.   We will refer to this chain as the \emph{simulated chain}.

A third Markov chain is constructed, with the corresponding box
covering shown in the bottom panel of Fig.\@~\ref{fig:D}.  This
uses a subset of the simulated float trajectories with $x_0$ chosen
to lie \emph{closest} to those of the observed floats.  This is
done to test the dependence of the TPT analysis on sampling.  We
will refer to this chain as the \emph{truncated chain}.

The source set $\mathcal A$ for TPT analysis is chosen to include
one box and to be located in the northernmost end of the Iceland
Basin, where ISOW enters the subpolar North Atlantic. The nominal
geographic location of the box is (16.37$^{\circ}$W, 62.55$^{\circ}$N),
which slightly varies depending on the Markov chain.  The target
set  $\mathcal B$  is placed along 49.25$^{\circ}$N, to frame export
paths of ISOW out of the subpolar North Atlantic.  This is the
southernmost and longest row of boxes for the three chains considered.

\begin{figure}[h!]
  \centering%
  \includegraphics[width=\columnwidth]{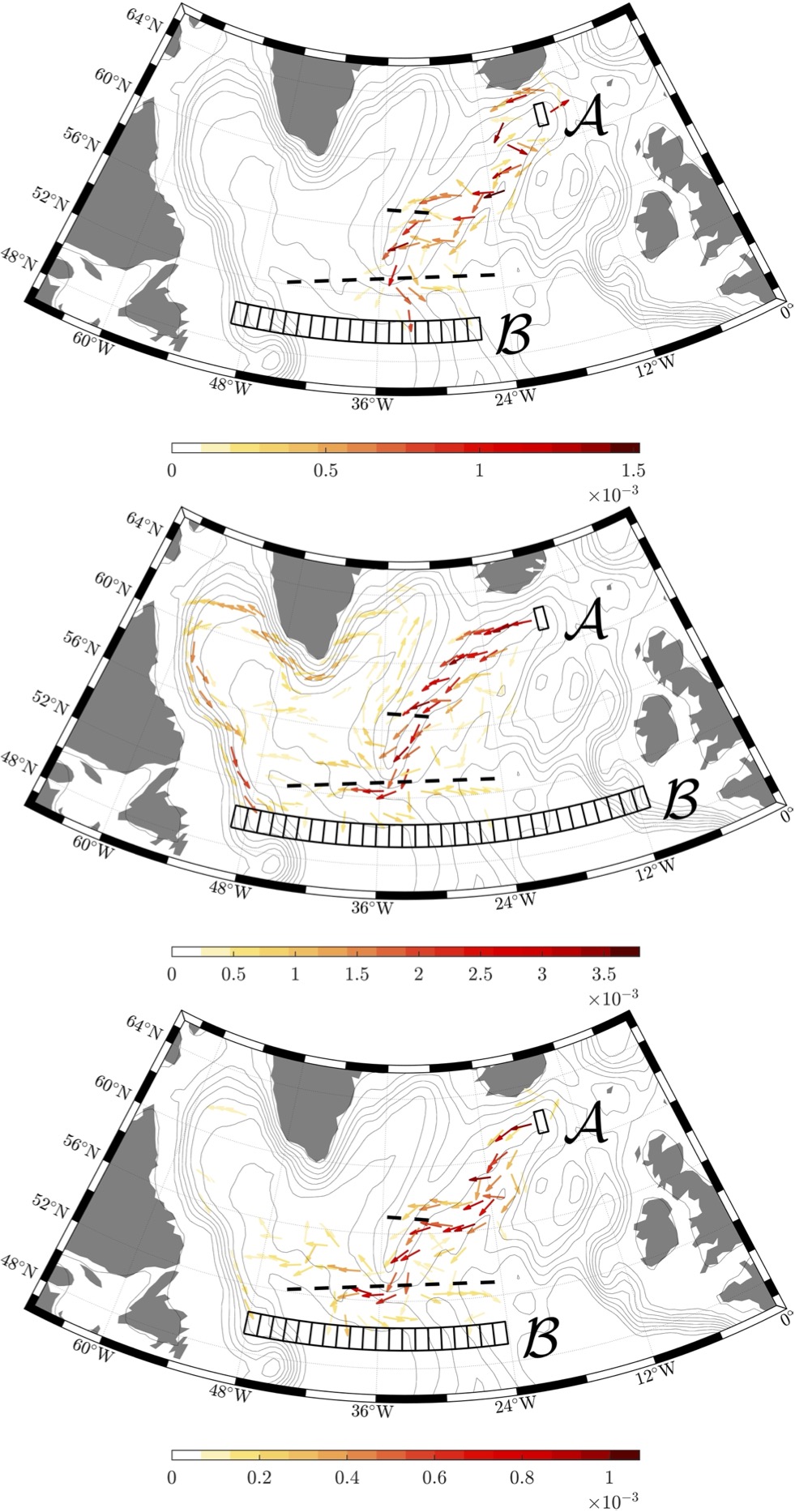}
  \caption{Reactive currents of ISOW as described by observed floats
  (top panel), simulated floats (middle panel), and simulated floats
  starting at observed float positions (bottom panel). The thin
  gray curves are the isobaths shown in Fig.\@~\ref{fig:subpolar}.
  The broken straight lines indicate the Bight Fracture Zone (northern line) 
  and the Charlie--Gibbs Fracture Zone (southern line).}
  \label{fig:f}%
\end{figure}

\section{TPT analysis of ISOW}

We begin by showing in Fig.\@~\ref{fig:f} the resulting reactive
currents.  To visualize them, we follow \citet{Helfmann-etal-20}
and for each box $B_i$ of the covering of $\mathcal D$, we estimate
the vector of the average direction and  magnitude of the reactive
current ($f^+$) to other boxes $B_j$, $j\neq i$. In agreement
with traditional abyssal circulation theory, the reactive currents
out of the source in the simulated chain (Fig.\@~\ref{fig:f}, middle
panel) organize into a DBC, which turns northward following the
bathymetry of the western flank of the Reykjanes Ridge, then continues
counter-clockwise around the Irminger Sea, and subsequently around
of the Labrador Sea after turning northward around the southern tip
of Greenland.  This happens before the reactive currents reach the
westernmost boxes of the target, in the Newfoundland Basin, which
are also reached by reactive currents that shortcut the DBC across
the Charlie--Gibbs Fracture Zone.  This is in stark contrast to
the reactive currents supported by the observed chain (Fig.\@~\ref{fig:f},
top panel), which do not reveal a DBC, but rather reach the target
east of the Mid-Atlantic Ridge.  The picture, however, is not too
different than that one drawn by the truncated chain (Fig.\@~\ref{fig:f},
bottom panel), suggesting that the absence of a well-defined DBC
in the observed chain might be due to insufficient sampling.

\begin{figure}[h!]
  \centering%
  \includegraphics[width=\columnwidth]{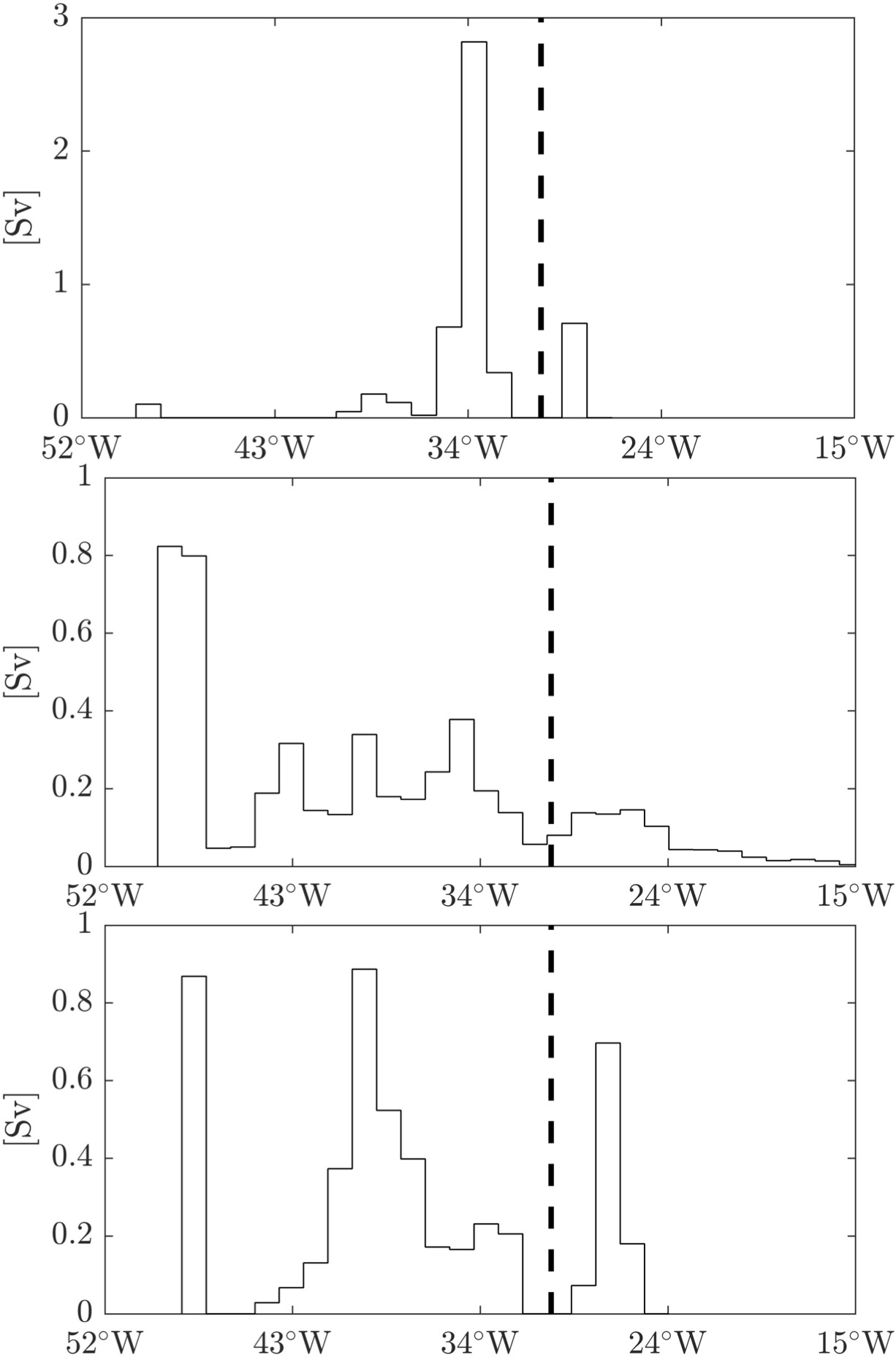}%
  \caption{Reactive rates, converted into volumetric fluxes, into
  each of the target boxes for the Markov chain constructed using observed
  floats (top panel), simulated floats (middle panel), and simulated
  floats starting at observed float positions (bottom panel). The
  dashed line indicates the longitude where the target
  intersects the Mid-Atlantic ridge.}
  \label{fig:k}%
\end{figure}

The reactive rates into each of the target boxes, \eqref{eq:decomp_rate},
offer, as expected, a picture consistent with the reactive currents.
These are depicted in Fig.\@~\ref{fig:k} as a function of longitude
along the target upon transforming them into volumetric flow rates.
This is done by first normalizing the reactive rates into each
target box by the total reactive rate into the target, and then
multiplying them by the volumetric flow rate of ISOW into the subpolar
North Atlantic inferred from in-situ observations (5\,Sv).  While
in the simulated chain (Fig.\@~\ref{fig:k}, middle panel) the volumetric
flow rate peaks in the western end of the target, in the observed chain
(Fig.\@~\ref{fig:k}, top panel) and truncated chain (Fig.\@~\ref{fig:k},
bottom panel) it also peaks east of the Mid-Atlantic Ridge.

\begin{figure}[h!]
  \centering%
  \includegraphics[width=\columnwidth]{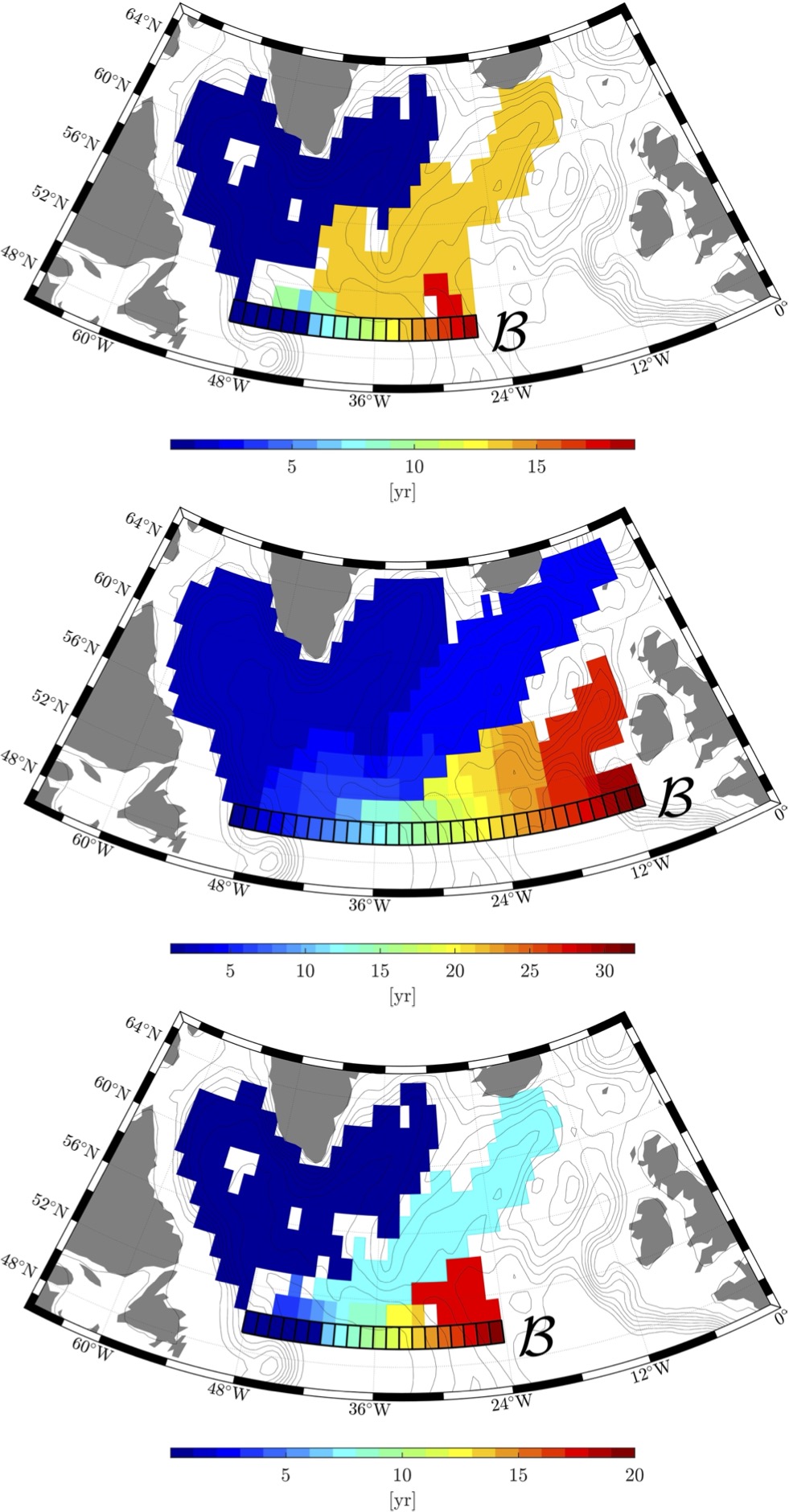}%
  \caption{Forward-committor-based dynamical geography revealing
  domains of influence for each box of the target with the provinces
  colored according to residence time for the observed (top panel),
  simulated (middle panel), and truncated (bottom panel) Markov
  chain. The gray curves are the isobaths shown in
  Fig.\@~\ref{fig:subpolar}.} 
  \label{fig:geo}%
\end{figure}

Finally, we present in Fig.\@~\ref{fig:geo} the domain of influence
of each box of the target. This is done by associating to each
domain box $B_i \in \mathcal D_N$ the most likely target box $B_j
\in \mathcal B$ to hit according to the probability in $B_i \in
\mathcal D_N$ to forward-commit to $B_j \in \mathcal B$ (and not
to any other box in $\mathcal B$).  This committor probability is
computed using \eqref{eq:q} for the plus sign with $P^+ = \tilde
P$,  $\mathcal A = \omega \cup (\mathcal B\setminus B_j)$, and
$\mathcal B$ replaced by $B_j \in \mathcal B$.  This way every box
of the covering of $\mathcal D$ gets assigned to a target box that
it will most likely reach, forming what \citet{Miron-etal-21-Chaos}
have dubbed a \defn{forward-committor-based dynamical geography}.
Each province $\mathcal P \subset \mathcal D_N$ shown in Fig.\
\ref{fig:geo}, i.e., each set of boxes that are most likely mapped
into a certain target box, is colored according to the \defn{expected
exit time}, $\smash{\frac{T}{1-\lambda_{\mathcal P}}}$, where
$\lambda_{\mathcal P}$ is the dominant eigenvalue of $\smash{\restr{\tilde
P}{\mathcal P,\mathcal P}}$ \citep[e.g.,][eq.\@~(10)]{Miron-etal-19-JPO},
which represents a measure of residence time in $\mathcal P$. The
geographies are consistent with the reactive currents in
Fig.\@~\ref{fig:f} and the volumetric volume rates in Fig.\@~\ref{fig:k}.
The observed chain (Fig.\@~\ref{fig:geo}, top panel) reveals a
marked partition, with the Irminger and Labrador Seas and Newfoundland
Basin predominantly forward committing to the westernmost end of
the target, and the Iceland Basin to a target box flanking the
Mid-Atlantic Ridge on the west. Moreover, the Irminger and Labrador
Seas and Newfoundland Basin have a shorter residence time (about 2
yr) compared to that of the Iceland Basin (about 14 yr). Altogether,
this suggests that the Irminger and Labrador Seas and Newfoundland
Basin are dynamically disconnected from the Iceland Basin.  This
is consistent with the absence of a DBC for the reactive currents
emerging from the northern Iceland Basin.  By contrast, in the
simulated chain (Fig.\ \ref{fig:geo}, middle panel) the Iceland
Basin is forward committed to the western side of the target and
has a residence time comparable to that of the Irminger and Labrador
Seas and Newfoundland Basin (2 to 5 yr, overall). This suggests a
dynamical connection of the Iceland Basin with the Irminger and
Labrador Seas and Newfoundland Basin, enabling, as a consequence,
reactive currents starting in the northern Iceland Basin to develop
a DBC.  But for the truncated chain (Fig.\@~\ref{fig:geo}, bottom
panel), the Iceland Basin presents a similar lack of communication
with the Irminger and Labrador Seas and Newfoundland Basin as in
the observed chain, highlighting the possibility that insufficient
sampling by floats may be masking the emergence of a DBC of ISOW.

\section{Testing the sampling sensitivity}

\begin{figure}[h!]
  \centering%
  \includegraphics[width=\columnwidth]{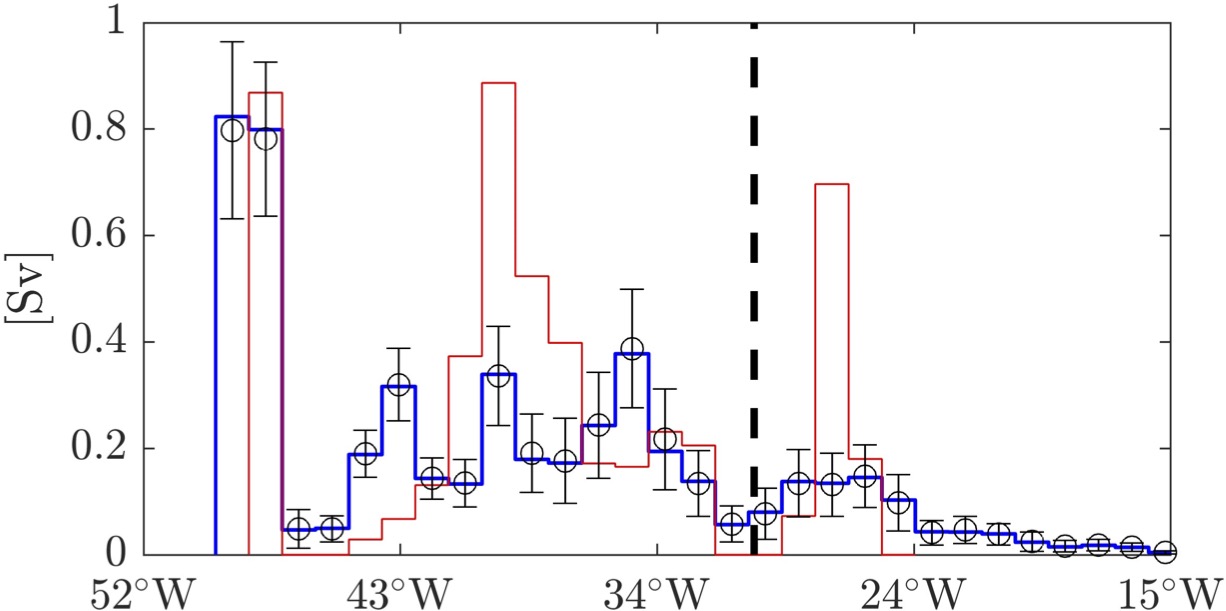}%
  \caption{As in Fig.\@~\ref{fig:k}, but for the simulated Markov
  chain after reducing uniformly at random the amount of trajectories
  involved its construction by 93\pct{ }, the circles represent
  averages over 100 uniform random reductions and the error bars
  correspond to one standard deviation about the mean values.  The
  blue curve is the result obtained using the full (i.e., with no
  trajectory reduction) simulated Markov chain (Fig.\@~\ref{fig:k},
  middle panel) while the red curve is that one obtained using the
  truncated chain (Fig.\@~\ref{fig:k}, bottom panel), which uses
  about 7\pct{ }of the total simulated trajectories, but starting
  at the observed float positions.}
  \label{fig:k-reducton}%
\end{figure}

The conclusion above on insufficient sampling deserves further
consideration.  In what follows we first test its robustness by
considering the results from reducing uniformly at random the number
of simulated trajectories $(x_0, x_T)$ involved in the construction
of the simulated Markov chain until it agrees with the amount in
the truncated chain.  Figure \ref{fig:k-reducton} shows the results
for the computation of the reactive rates into the target boxes at
the southern edge of the domain.  The circles are averages over 100
uniform random reductions of the trajectories by 93\pct.  These
averages are accompanied by one-standard-deviation error bars.
Overlaid in blue is the reactive rate obtained using the simulated
Markov chain computed using all trajectories at disposal.  As can
be seen, this result is surprisingly very robust, given that only
7\pct{ }of the trajectories are considered.  By contrast, the
reactive rates obtained using the truncated Markov chain (red curve
in Fig.~\ref{fig:k-reducton}) are quite different and lie nearly
everywhere outside of the variability of the rates of the uniformly
reduced Markov chain, even though it uses approximately the same
amount of trajectories.

\begin{figure}[h!]
  \centering%
  \includegraphics[width=\columnwidth]{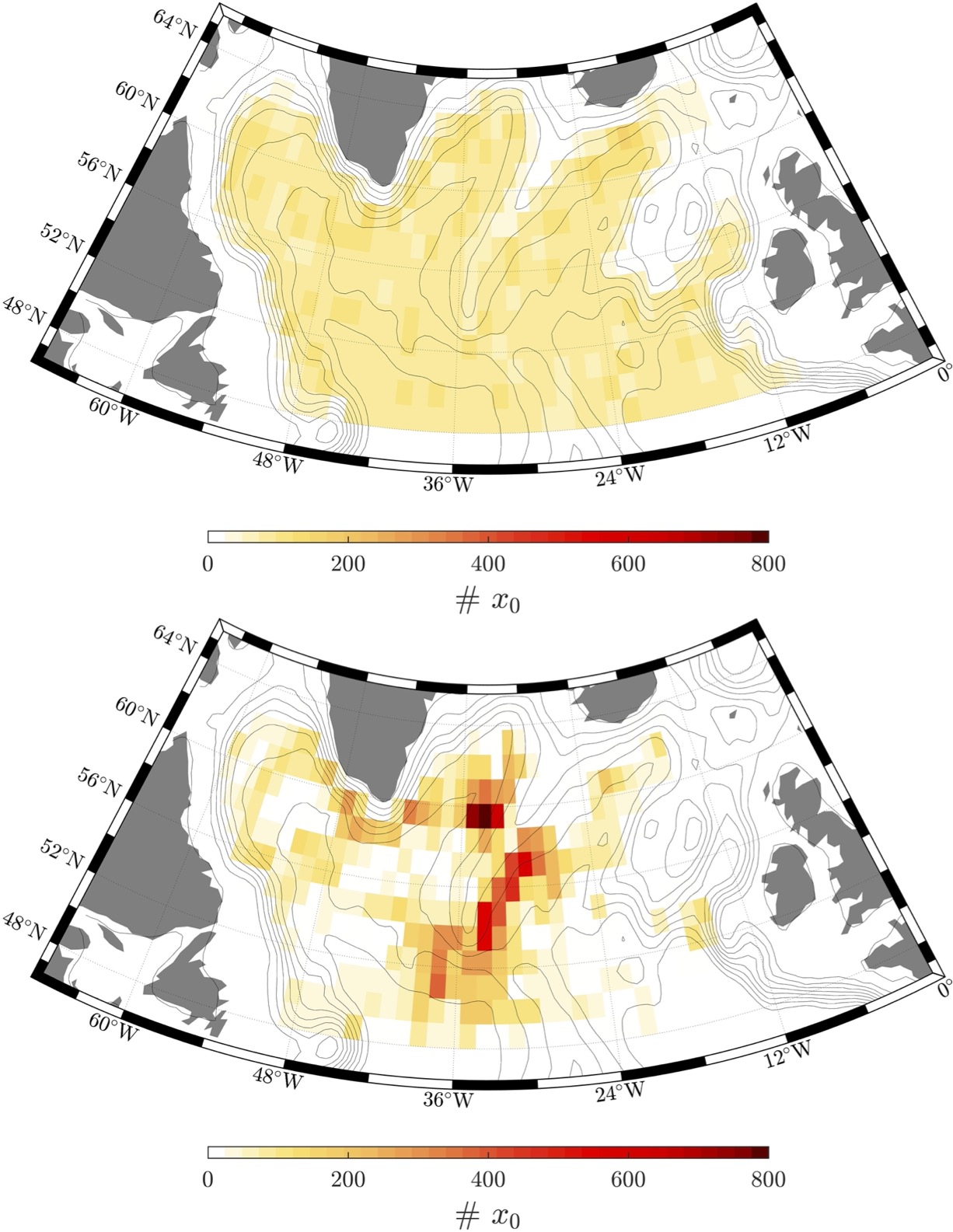}%
  \caption{(top panel) Histogram of simulated float initial positions,
  $x_0$, keeping only 7\pct{ }of them, uniformly at random. The
  thin gray curves are the isobaths shown in Fig.\@~\ref{fig:subpolar}
  (bottom panel). As in the top panel, but with $x_0$ lying closest
  to those of the observed float trajectories.}
  \label{fig:x0}%
\end{figure}

The reason for this difference in the results between the uniformly
reduced chain and the truncated chain is that in the former case
the trajectory sampling is uniform, and in the latter case the
sampling is nonuniform, as it is shown in Fig.\@~\ref{fig:x0}.  The
top panel of this figure shows a histogram of the simulated float
initial positions $x_0$ keeping only 7\pct{ }of them, uniformly at
random.  This contrasts with the distribution in the bottom panel,
which shows the initial float positions $x_0$ used for constructing
the truncated Markov chain.  Since the initial positions, in this
case, lie closest to those of the observed floats, the lack of
homogeneity in the distribution is dictated by the sampling strategy
taken during the OSNAP experiment, with most of the floats deployed
on the western and eastern flanks of the Reykjanes Ridge. From this
experiment, we can also suspect that the TPT results obtained from
the observed chain are biased due to the nonuniform sampling of the
floats.

\begin{figure}[h!]
  \centering%
  \includegraphics[width=\columnwidth]{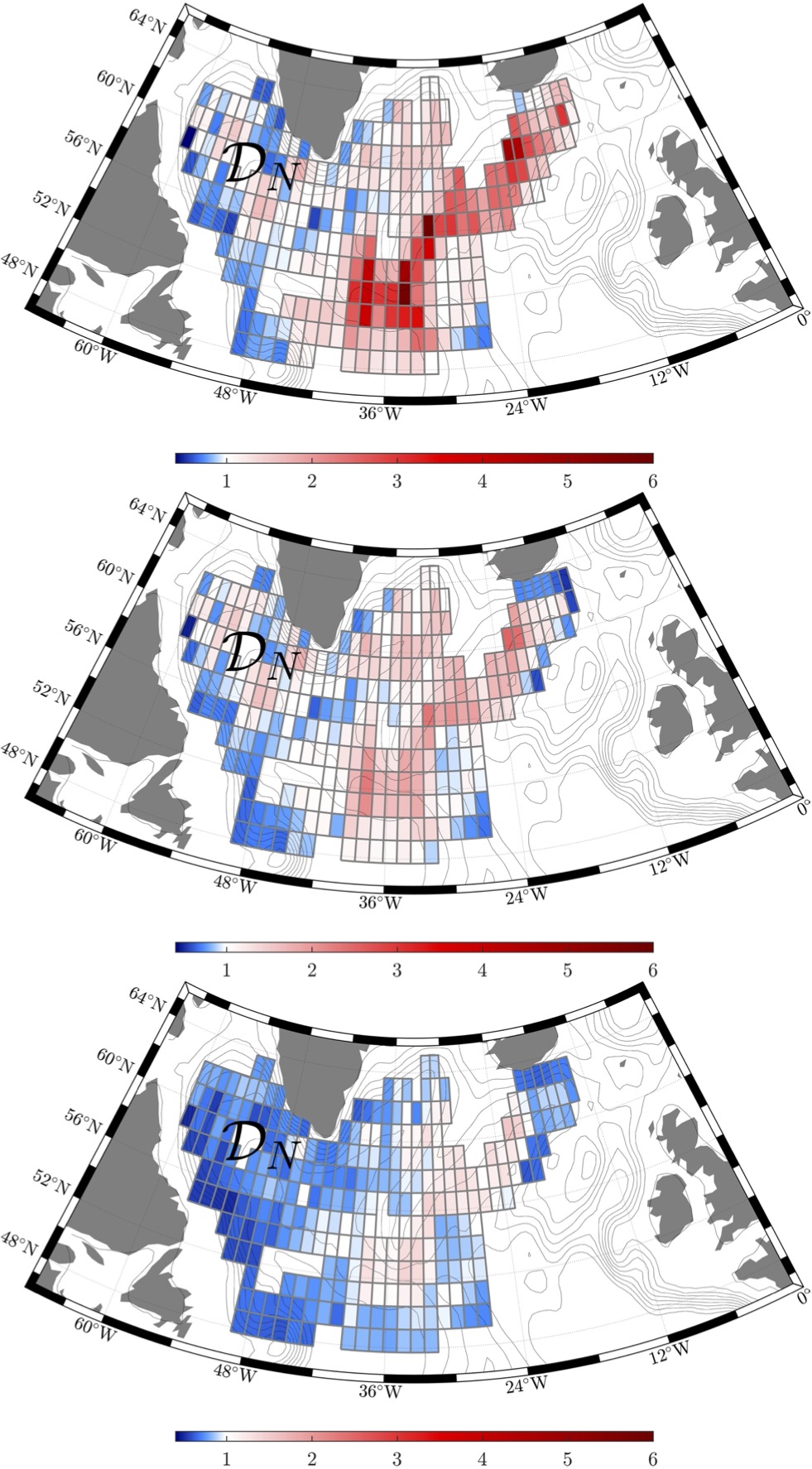}% 
  \caption{Averaged signal-to-noise ratio of the transition probability
  between each covering box and neighboring boxes for the observed
  Markov chain as estimated from 100 realizations thereof constructed
  using 75 (top panel), 50 (middle panel), and 25 (bottom panel)
  percent of the available float trajectories, chosen uniformly at
  random.  The thin gray curves are the isobaths shown in
  Fig.\@~\ref{fig:subpolar}.}
  \label{fig:pvar}%
\end{figure}

Further support to the need of better sampling for reliable TPT
computations is given by studying different realizations of the
observed Markov chain constructed with a uniformly at random reduced
set of trajectories $(x_0, x_T)$. Over these different realizations,
we consider the signal-to-noise ratio S/R (mean over standard
deviation, also the inverse of the coefficient of variation) as a
dimensionless and comparable measure of variability and uncertainty
in the data. When S/N is smaller than 1, the standard deviation is
larger than the mean, and thus there is large uncertainty and
variability in the estimation of the observed chain. In particular,
we compute the S/R for each transition probability $P_{ij}$ from
box $B_i$ to box $B_j$ over the different realizations of uniformly
at random reduced observed chains. We denote the S/N of each entry
$P_{ij}$ by $S_{ij}$.  In order to show the resulting ratios on the
domain and since the matrix $S$ inherits the sparsity of $P$ and
is often zero when $B_i$ and $B_j$ are not neighbors, we average
$S_{ij}$ for each outbound box $B_i$ over the set of neighboring
boxes. The vector of average ratios is denoted by $\mathbf s \in
\mathbb R^{1\times N}$. The ratio $s_i$ of box $B_i$ tells us the
amount of uncertainty in the estimation of transition probabilities
out of box $B_i$.  A reliable TPT analysis requires a transition
matrix $P$ with $\mathbf s$ componentwise (sufficiently) greater
than 1. The top, middle, and bottom panels of Fig.\@~\ref{fig:pvar}
show estimates of $\mathbf s$ for 100 realizations of $P$ obtained
by uniformly keeping at random 75, 50, and 25\pct, respectively,
of the available float trajectories.  The ``noise'' consistently
masks the ``signal'' in the majority of the domain.  A consistent
exception is the eastern flank of the Reykjanes Ridge.  This provides
reason to distrust the results from the TPT analysis of the observed
Markov chain, demanding an improvement of the float data set with
more sampling, especially in the western regions where $\mathbf s$
is componentwise smaller than $1$.

\section{Summary and conclusions}

The equatorward export of Iceland--Scotland Overflow Water (ISOW)
from the subpolar North Atlantic is a problem that has stirred much
debate in recent years. The ISOW is the lighter of the two overflow
components of the North Atlantic Deep Water (NADW), which forms the
upper limb of Atlantic Meridional Overturning Circulation (AMOC)
that flows southward.  The AMOC strength and the impacts on the
planet's climate regulation depend on the rates of formation of
NADW, and ISOW in particular.  Recent analyses of observed deep
float trajectories have led investigators to conclude that the
traditional abyssal circulation should be revised.  Such a theory
postulates that ISOW should steadily flow equatorward along a
well-organized deep boundary current (DBC) around the subpolar North
Atlantic.  Unlike those analyses, which involved direct inspection
of observed trajectories and the construction of probability
distributions from long trajectories integrated using numerically
generated velocities, which typically lead to a picture difficult
to interpret, here we have applied Transition Path Theory (TPT).
TPT is designed to rigorously characterize the ensembles of
trajectories that directly connect a source with a target.  As such,
TPT is very well suited to tackle the ISOW problem as details of
individual connecting trajectories are averaged out and instead the
average dominant transport channels are found.

TPT was applied on three different time-homogeneous Markov chains
on boxes that covered the subpolar North Atlantic, which are useful
for framing long-term ISOW motion asymptotics. One was constructed
using a high number of simulated trajectories homogeneously covering
the flow domain. The other two chains used much fewer trajectories
that heterogeneously covered the domain. The trajectories in the
latter two chains were observed trajectories or simulated trajectories
subsampled at the observed frequency.  While the densely sampled
chain supported a well-defined DBC, the sparsely sampled chains did
not, independent of whether observed or simulated trajectories were
involved.  By studying the sensitivity of the results to sampling,
we conclude that heterogeneous and insufficient sampling might be
behind the current debate around the validity of the traditional
abyssal circulation theory.

While the required sampling density to unveil, or not, a DBC of
ISOW seems beyond the capability of observational platforms at
present, the TPT analysis, or any analysis aimed at investigating
long-time asymptotics and connectivity, might benefit from a
subsurface float dataset enlarged with extra floats deployed in the
northern Iceland Basin, where the ISOW enters the subpolar North
Atlantic, and in the western flank of the  Reykjanes Ridge.  The
expectation is that trajectories initialized there will reveal any
missed communication between the basins west and east of the
Reykjanes/Mid-Atlantic Ridge, and also within these basins.  The
western side, more specifically the Irminger and Labrador Seas, and
the Newfoundland Basin, is characterized by transition probabilities
with a low signal-to-noise ratio and hence high uncertainty.  In
turn, the eastern side has a large region, the West European Basin,
that is not visited by float trajectories at all.  The lack of
sampling there attempts against the significance of any transition
export paths flanking the Mid-Atlantic Ridge on the east.

\acknowledgments We thank Xiaobiao Xu for extracting and making
available to us the HYCOM model velocity data, Susan Lozier for the
benefit of many discussions on overflow dynamics, Alexander Sikorski
for discussions on testing data sensitivity, and to P\'eter Koltai
for suggestions about the construction of the Markov chain for ISOW.
This work was supported by NSF grant OCE1851097.

\datastatement The RAFOS float data are distributed by the National
Oceanic and Atmospheric Administration's (NOAA's) Atlantic Oceanographic
and Meteorological Laboratory (AOML) through the subsurface data
sets website at https://www.aoml.noaa.gov/phod/float\_traj/.  The
trajectories of the Argo floats at their parking level are available
in near-real-time at http://apdrc.soest.hawaii.edu/projects/yomaha.
The HYCOM output may be available from request to Xiaobiao Xu.

\bibliographystyle{ametsocV6} 
\bibliography{fot}

\begin{thebibliography}{20}
\providecommand{\natexlab}[1]{#1}
\providecommand{\url}[1]{\texttt{#1}}
\renewcommand{\UrlFont}{\rmfamily}
\providecommand{\urlprefix}{URL }
\expandafter\ifx\csname urlstyle\endcsname\relax
  \providecommand{\doi}[1]{https://doi.org/\discretionary{}{}{}#1}\else
  \providecommand{\doi}{https://doi.org/\discretionary{}{}{}\begingroup
  \urlstyle{rm}\Url}\fi
\providecommand{\eprint}[2][]{\url{#2}}

\bibitem[{Argo(2000)}]{Argo-00}
Argo, 2000: {Argo float data and metadata from Global Data Assembly Centre
  (Argo GDAC)}. SEANOE. http://doi.org/10.17882/42182.

\bibitem[{Buckley and Marshall(2016)Buckley, and
  Marshall}]{Buckley-Marshall-16}
Buckley, M.~W., and J.~Marshall, 2016: {Observations, inferences, and
  mechanisms of Atlantic Meridional Overturning Circulation variability: A
  review}. \textit{Rev. Geophys.}, \textbf{54}, 10.1002/2015RG000\,493.

\bibitem[{Daniault et~al.(2016)}]{Daniault-etal-16}
Daniault, N., and Coauthors, 2016: {The northern North Atlantic Ocean mean
  circulation in the early 21st century}. \textit{Progress in Oceanography},
  \textbf{146}, 142--158.

\bibitem[{{E} and {Vanden-Eijnden}(2006){E}, and
  {Vanden-Eijnden}}]{E-vandenEijnden-06}
{E}, W., and E.~{Vanden-Eijnden}, 2006: Towards a theory of transition paths.
  \textit{J. Stat. Phys.}, \textbf{123}, 503--623.

\bibitem[{Helfmann et~al.(2020)Helfmann, Borrell, Sch\"utte,, and
  Koltai}]{Helfmann-etal-20}
Helfmann, L., E.~R. Borrell, C.~Sch\"utte, and P.~Koltai, 2020: Extending
  transition path theory: {P}eriodically driven and finite-time dynamics.
  \textit{J. Nonlinear Sci.}, \textbf{30}, 3321--3366.

\bibitem[{Johns et~al.(2021)Johns, Devana, Houk,, and Zou}]{Johns-etal-21}
Johns, W.~E., M.~Devana, A.~Houk, and S.~Zou, 2021: {Moored Observations of the
  Iceland-Scotland Overflow Plume Along the Eastern Flank of the Reykjanes
  Ridge}. \textit{Journal of Geophysical Research}, \textbf{126},
  e2021JC017\,524.

\bibitem[{Koltai(2010)}]{Koltai-10}
Koltai, P., 2010: Efficient approximation methods for the global long-term
  behavior of dynamical systems -- theory, algorithms and examples. Ph.D.
  thesis, Technical University of Munich.

\bibitem[{Lozier et~al.(2017)}]{Lozier-etal-17}
Lozier, S.~M., and Coauthors, 2017: {Overturning in the Subpolar North Atlantic
  Program: A New International Ocean Observing System}. \textit{Bulletin of the
  American Meteorological Society}, \textbf{98}, 737--752.

\bibitem[{Metzner et~al.(2009)Metzner, Sch\"utte,, and
  Vanden-Eijnden}]{Metzner-etal-09}
Metzner, P., C.~Sch\"utte, and E.~Vanden-Eijnden, 2009: Transition path theory
  for {M}arkov jump processes. \textit{Multiscale Modeling \& Simulation},
  \textbf{7}, 1192--1219.

\bibitem[{Miron et~al.(2021)Miron, Beron-Vera, Helfmann,, and
  Koltai}]{Miron-etal-21-Chaos}
Miron, P., F.~J. Beron-Vera, L.~Helfmann, and P.~Koltai, 2021: Transition paths
  of marine debris and the stability of the garbage patches. \textit{Chaos},
  \textbf{31}, 033\,101.

\bibitem[{Miron et~al.(2022)Miron, Beron-Vera,, and Olascoaga}]{Miron-etal-22}
Miron, P., F.~J. Beron-Vera, and M.~J. Olascoaga, 2022: {Transition paths of
  North Atlantic Deep Water}. \textit{J. Atmos. Oce. Tech.}, \textbf{39},
  959–971.

\bibitem[{Miron et~al.(2019{\natexlab{a}})Miron, Beron-Vera, Olascoaga,
  Froyland, P\'erez-Brunius,, and Sheinbaum}]{Miron-etal-19-JPO}
Miron, P., F.~J. Beron-Vera, M.~J. Olascoaga, G.~Froyland, P.~P\'erez-Brunius,
  and J.~Sheinbaum, 2019{\natexlab{a}}: {Lagrangian geography of the deep Gulf
  of Mexico}. \textit{J. Phys. Oceanogr.}, \textbf{49}, 269--290.

\bibitem[{Miron et~al.(2019{\natexlab{b}})Miron, Beron-Vera, Olascoaga,, and
  Koltai}]{Miron-etal-19-Chaos}
Miron, P., F.~J. Beron-Vera, M.~J. Olascoaga, and P.~Koltai,
  2019{\natexlab{b}}: {Markov-chain-inspired search for MH370}. \textit{Chaos:
  An Interdisciplinary Journal of Nonlinear Science}, \textbf{29}, 041\,105.

\bibitem[{Norris(1998)}]{Norris-98}
Norris, J., 1998: \textit{Markov Chains}. Cambridge University Press.

\bibitem[{Rossby et~al.(1986)Rossby, Dorson,, and Fontaine}]{Rossby-etal-86}
Rossby, T., D.~Dorson, and J.~Fontaine, 1986: {The RAFOS system}. \textit{J.
  Atmos. Ocean. Technol.}, \textbf{3}, 672--679.

\bibitem[{Steele et~al.(1962)Steele, Barrett,, and
  Worthington}]{Steele-etal-62}
Steele, J., J.~Barrett, and L.~Worthington, 1962: {Deep currents south of
  Iceland}. \textit{Deep Sea Research and Oceanographic Abstracts}, \textbf{9},
  465--474, \doi{https://doi.org/10.1016/0011-7471(62)90097-9}.

\bibitem[{Stommel(1958)}]{Stommel-58}
Stommel, H., 1958: The abyssal circulation. \textit{Deep Sea Res.}, \textbf{5},
  80?82.

\bibitem[{Tarjan(1972)}]{Tarjan-72}
Tarjan, R., 1972: Depth-first search and linear graph algorithms. \textit{SIAM
  J. Comput.}, \textbf{1}, 146--160.

\bibitem[{Xu et~al.(2013)Xu, Hurlburt, Schmitz~Jr., Zantopp, Fischer,, and
  Hogan}]{Xu-etal-13}
Xu, X., H.~E. Hurlburt, W.~J. Schmitz~Jr., R.~Zantopp, J.~Fischer, and P.~J.
  Hogan, 2013: {On the currents and transports connected with the atlantic
  meridional overturning circulation in the subpolar North Atlantic}.
  \textit{Journal of Geophysical Research: Oceans}, \textbf{118}, 502--516.

\bibitem[{Zou et~al.(2020)Zou, Bower, Furey,, and Lozier}]{Zou-etal-20}
Zou, S., A.~Bower, H.~Furey, and S.~Lozier, 2020: {Redrawing the
  Iceland--Scotland Overflow Water pathways in the North Atlantic}.
  \textit{Nat. Commun.}, \textbf{11}, 1890.

\end{thebibliography}

\end{document}